\documentclass[letterpaper]{article} 
\usepackage{aaai2026}  
\usepackage{times}  
\usepackage{helvet}  
\usepackage{courier}  
\usepackage[hyphens]{url}  
\usepackage{graphicx} 
\usepackage{natbib}  
\usepackage{caption} 
\usepackage{algorithm}
\usepackage{algorithmic}

\usepackage{newfloat}
\usepackage{listings}
\DeclareCaptionStyle{ruled}{labelfont=normalfont,labelsep=colon,strut=off} 
\floatstyle{ruled}
\newfloat{listing}{tb}{lst}{}
\floatname{listing}{Listing}
\title{On Evaluating the Adversarial Robustness of Foundation Models for Multimodal Entity Linking}

\author{
    Fang Wang\textsuperscript{\rm 1},
    Yongjie Wang\textsuperscript{\rm 2},
    Zonghao Yang\textsuperscript{\rm 3},
    Minghao Hu\textsuperscript{\rm 4},
    Xiaoying Bai\textsuperscript{\rm 4}\thanks{Correspond author.}
}
\affiliations{
    \textsuperscript{\rm 1}Peking University\\
    \textsuperscript{\rm 2}Hangzhou Zhiyuan Research Institute Co., Ltd.\\
    \textsuperscript{\rm 3}China Research and Development Academy of Machinery Equipment\\
    \textsuperscript{\rm 4}Advanced Institute of Big Data,Beijing\\
    fangwang@stu.pku.edu.cn, \{wangyhongjie, huminghao16\}@gmail.com, yangzonghao1024@163.com, baixy@aibd.ac.cn
}

\usepackage{bibentry}

\usepackage{amssymb} 
\usepackage{bm}
\usepackage{bbding} 
\usepackage{multirow} 
\usepackage{color} 
\usepackage{hhline}
\usepackage{booktabs} 
\usepackage{boldline}
\usepackage{amsmath} 
\usepackage{array} 
\usepackage{enumerate} 
\usepackage{pdfpages}
\usepackage{subfigure}
\usepackage{tabularx}
\usepackage{makecell}
\usepackage{rotating}

\begin{document}

\maketitle

\begin{abstract}
The explosive growth of multimodal data has driven the rapid development of multimodal entity linking (MEL) models. However, existing studies have not systematically investigated the impact of visual adversarial attacks on MEL models. We conduct the first comprehensive evaluation of the robustness of mainstream MEL models under different adversarial attack scenarios, covering two core tasks: Image-to-Text (I2T) and Image+Text-to-Text (IT2T). Experimental results show that current MEL models generally lack sufficient robustness against visual perturbations. Interestingly, contextual semantic information in input can partially mitigate the impact of adversarial perturbations. Based on this insight, we propose an LLM and Retrieval-Augmented Entity Linking (LLM-RetLink), which significantly improves the model's anti-interference ability through a two-stage process: first, extracting initial entity descriptions using large vision models (LVMs), and then dynamically generating candidate descriptive sentences via web-based retrieval. Experiments on five datasets demonstrate that LLM-RetLink improves the accuracy of MEL by 0.4\%-35.7\%, especially showing significant advantages under adversarial conditions. This research highlights a previously unexplored facet of MEL robustness, constructs and releases the first MEL adversarial example dataset~\footnote{https://anonymous.4open.science/r/MEL-Robustness-90A5}, and sets the stage for future work aimed at strengthening the resilience of multimodal systems in adversarial environments.
\end{abstract}
\section{Introduction}
With the explosive growth of multimodal information on the Internet, Multimodal Entity Linking has emerged as a crucial bridge connecting visual content with structured knowledge. It plays a central role in cross-modal information understanding and has been widely applied in various downstream tasks such as knowledge-enhanced question \& answering~\cite{hu2024ket}, image-text retrieval~\cite{huang2024cross}, and open-domain entity alignment~\cite{qiu2024entity6k}. In recent years, driven by advances in pretrained large language models, large vision models, and retrieval-augmented mechanisms, MEL has made significant progress in cross-modal representation learning, candidate generation, and entity disambiguation strategies~\cite{liang2024crossformer,pourreza2024chase,pons2024knowledge}. These developments have enabled MEL to consistently achieve state-of-the-art performance on multiple standard benchmark datasets, gradually establishing MEL as a key paradigm for achieving general image-text semantic understanding.

 \begin{figure}[!t]
    \centering
    \includegraphics[width=\columnwidth]{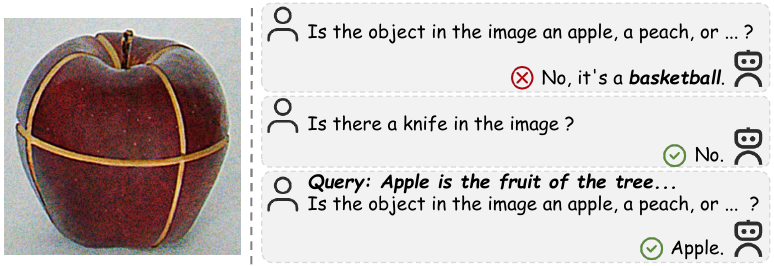}
    \caption{QA pairs for LLaVA given an adversarial image. In the first two QA pairs, LLaVA provides entirely incorrect answers. However, it correctly responds to subsequent questions that are unrelated to the attacked object (apple). Notably, the contrast between the first and last QA pairs highlights that LLaVA can answers the question correctly once additional context is introduced. These observations support key findings in this paper. Source: $M^3EL$~\cite{WF-M3EL}}
    \label{fig:llava_response}
\end{figure}

Despite recent advances in MEL, the impact of adversarial examples remains underexplored, particularly in terms of systematic evaluation against visual adversarial perturbations. Most existing MEL methods assume that visual inputs are clean and accurate, while in real-world scenarios, images are often subject to noise or even malicious attacks~\cite{chi2024adversarial}. Meanwhile, adversarial attacks have demonstrated strong deceptive capabilities in other multimodal tasks such as image classification and visual question answering~\cite{archana2024deep, sima2024drivelm}. However, there is still a lack of systematic investigation of the vulnerability of MEL models under such perturbations. This gap hinders the reliable deployment of MEL models in complex real-world environments.

From a practical perspective, MEL typically takes both images and text as inputs, with the overall processing pipeline involving multiple functional components such as a visual encoder, a text encoder, and an entity disambiguation module. This architectural complexity raises a critical question. Can MEL models maintain stable linking performance when adversarial perturbations are applied solely to the visual input? In other words, are localized perturbations sufficient to disrupt the overall cross-modal alignment structure, thereby affecting the final entity recognition and disambiguation outcomes? Furthermore, as a source of semantic context, can the textual modality help mitigate the semantic shift caused by visual perturbations and thus enhance the MEL's robustness? These important questions remain largely unexplored in existing methods and call for systematic investigation and empirical validation.

To this end, we conducted a comprehensive analysis of the robustness of current MEL models under various adversarial attacks, tasks, and datasets. We selected five representative datasets and focus on two core tasks, Image-to-Text Entity Linking (I2T-EL) and Image+Text-to-Text Entity Linking (IT2T-EL), to systematically evaluate model performance under visual adversarial perturbations. Experimental results reveal that MEL models exhibit pronounced vulnerability when textual context is absent. In contrast, incorporating textual information alongside visual input significantly improves the robustness of the model.

Overall, the contributions of our work can be summarized as follows:

\begin{itemize}
    \item We perform a comprehensive analysis of the robustness of current MEL methods against various adversarial attacks, tasks, and datasets.
    \item We release a visual adversarial dataset built on five common MEL benchmarks.
    \item We propose LLM-RetLink, which integrates the semantic understanding of LLMs with retrieval-based context enhancement. 
\end{itemize}


\section{Related Work}
\noindent \textbf{Multi-modal Entity Linking \quad}
The pioneering study by Moon~\cite{moon-etal-2018-multimodal-named}first demonstrated the effectiveness of visual information in disambiguating short-text mentions on social networks, proposing a zero-shot learning framework that integrates textual, visual, and lexical features. Subsequently, Adjali~\cite{adjali-etal-2020-building}introduced an automated MEL dataset construction method based on Twitter data. Gan~\cite{gan2021multimodal} formulated the alignment between textual and visual mentions as a bipartite graph matching problem. Wang~\cite{wang2022wikidiverse} used a multimodal co-attention mechanism to hierarchically extract jointly attended textual and visual features. Zhang~\cite{ZhangH22} proposed an IMLN to make full use of information in terms of mentions and knowledge. Xing~\cite{XingZWLZD23} introduced DRIN to capture fine-grained relationships between entities and mentions. Luo~\cite{LuoXWZXC23} proposed the MIMIC, which is designed to comprehensively model both abbreviated textual context and implicit visual cues, enabling more robust multimodal representation learning. Long~\cite{LongZMZZ24} introduced GELR to incorporate a knowledge retriever to enhance the visual information of the entity. Song~\cite{SongZ00LMW24} and Zhao~\cite{zhao2025me3a} implemented queries with multimodal data and addressed semantic gaps using cross-modal enhancers between text and image information.

\noindent \textbf{Adversarial attacks \quad}
Adversarial attacks refer to the addition of imperceptible perturbations to input, which mislead neural networks into making incorrect predictions. Based on the amount of information accessible to the attacker, these attacks can be categorized into three types: white-box attacks~\cite{Szegedyintrig, carlini2017, obfuscated-gradients}, where the attacker has full access to the model architecture and parameters; black-box attacks~\cite{Papernot_blackbox, Su2017OnePA}, where only input-output queries are available and internal details remain hidden; and gray-box attacks, where the attacker has partial knowledge of the model, such as its architecture.

\noindent \textbf{Adversarial Examples \quad}
Neural network models have been shown to be highly vulnerable to adversarial examples in Computer Vision tasks~\cite{carlini2017, madry2018towards, pmlr-v119-croce20b}. These adversarial examples are typically crafted by applying subtle and carefully designed perturbations to input data, which remain visually indistinguishable from the original but are sufficient to cause incorrect predictions. We evaluated the effectiveness of several well-known adversarial attack methods in manipulating model outputs, including Projected Gradient Descent (PGD)~\cite{madry2018towards}, Auto-PGD (APGD)~\cite{pmlr-v119-croce20b}, and the Carlini \& Wagner (CW) attacks~\cite{carlini2017}. Studying these attacks is essential for gaining a deeper understanding of model robustness and vulnerabilities, and provides theoretical foundations for developing effective defense mechanisms and assessing adversarial risks in ML systems.

\section{Method}
\subsection{Problem Formulation}
We consider gradient-based white-box adversarial attacks. Given an input-label pair $(x, y)$ and a classifier $f$, the objective is to find a perturbation $\delta$ such that  the prediction of the perturbed input is incorrect:

\begin{equation}
f(x + \delta) \ne y
\end{equation}

\textbf{PGD} seeks to maximize the classification loss while constraining the perturbation within an $L_\infty$ ball:

\begin{equation}
\max_{\|\delta\|_\infty \le \epsilon} \mathcal{L}(f(x + \delta), y) 
\end{equation}

where $\mathcal{L}$ denotes the cross-entropy loss, and $\epsilon$ controls the attack strength.

\textbf{CW} attack formulates the problem as follows:

\begin{equation}
\max_{\delta} \|\delta\|_p + c \cdot g(x + \delta), \quad x + \delta \in [0, 1]^n
\end{equation}

where $c>0$ is a balancing hyperparameter and the function $g$ is defined as:

\begin{equation}
g(x + \delta) = \max\left(f(x + \delta)_y - \max_{i \ne y} f(x + \delta)_i, -\kappa\right)
\end{equation}

where, $\kappa$ is the confidence margin controlling the strength of misclassification.

\begin{table}[!t]
\small
\centering
\begin{tabular}{cccccccc}
\toprule
Type                    & Method & Step & Size & $\epsilon$ & Dist. & $c$ & $\kappa$ \\ \midrule
\multirow{3}{*}{Normal} & PGD    & 20    & 2/255    & 8/255 & $L_{\infty}$     & - & - \\
                        & APGD   & 20    & -        & 8/255 & $L_{\infty}$     & - & - \\
                        & CW     & 50    & 0.01     & -     & $L_{2}$          & 20 & 0 \\ \midrule
\multirow{3}{*}{Strong} & PGD    & 40    & 2/225    & 0.2   & $L_{\infty}$     & - & - \\ 
                        & APGD   & 40    & -        & 0.2   & $L_{\infty}$     & - & - \\
                        & CW     & 75    & 0.05     & -     & $L_{2}$          & 100 & 0 \\
\bottomrule
\end{tabular}
\caption{Parameters for the attacks under Normal and Strong settings. Dist. refers to distance measure, $c$ and $\kappa$ refers to the constraint and confidence in ~\cite{cui2024robustness}.}
\label{tab: attack_parameters}
\end{table}

\begin{figure}[!b]
    \centering
    \includegraphics[width=0.75\columnwidth]{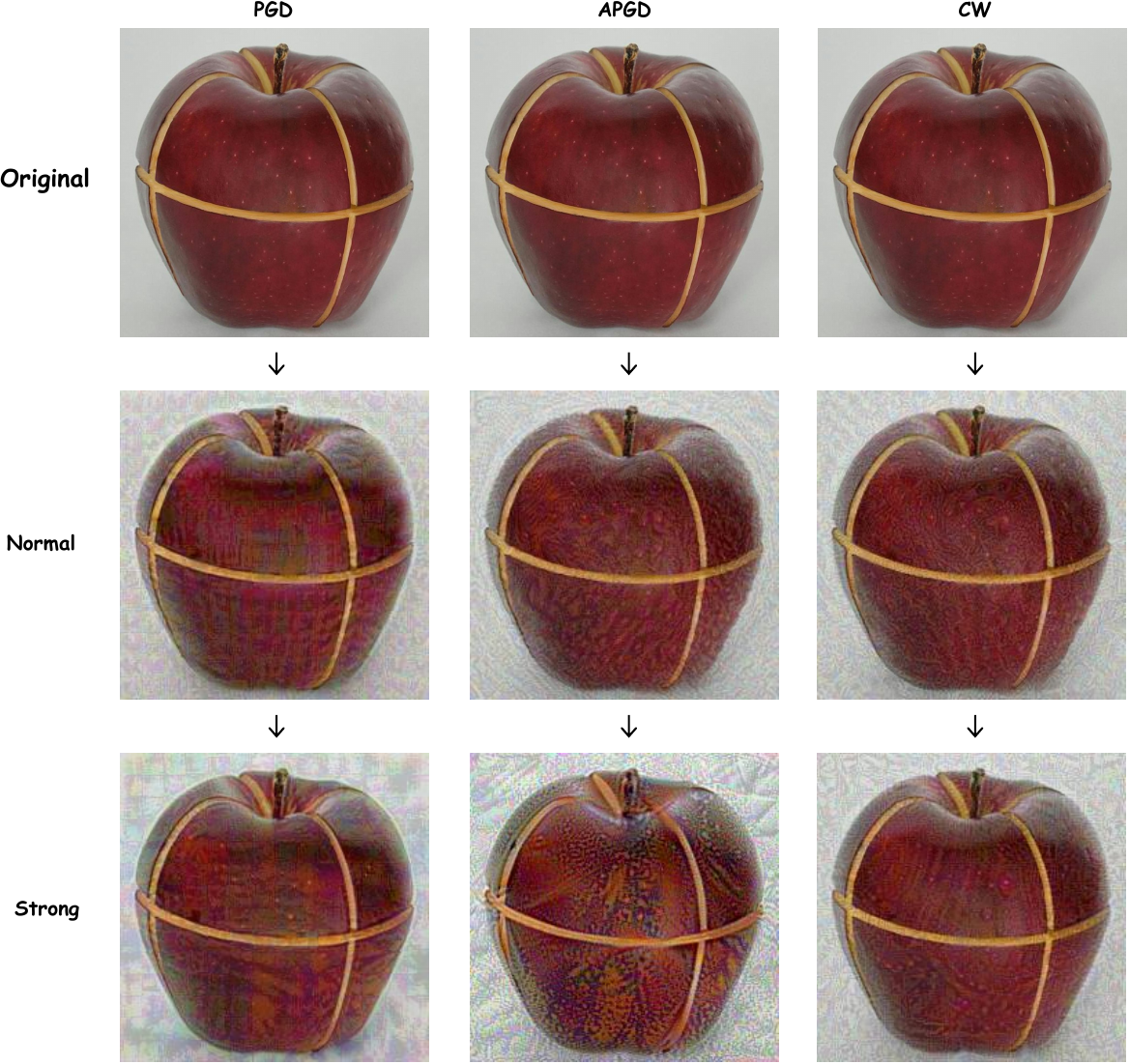}
    \caption{The samples of adversarial image, generated by PGD, APGD and CW, under Normal and Strong attack parameter settings. Image source: $M^3EL$.}
    \label{fig: adversarial_samples}
\end{figure}

\subsection{Attacks \& Adversarial generation}
We adopt PGD and CW as two representative gradient-based attacks, and include APGD as a variant of PGD. For each method, we consider two configurations, normal and strong, distinguished by the perceptual visibility of the perturbations. Detailed parameters are summarized in Table~\ref{tab: attack_parameters}.

\noindent \textbf{Normal \quad}
We follow previous works~\cite{cui2024robustness}, setting the constraint for CW to 20 and the $\epsilon=8/255$ for PGD / APGD. For brevity, we denote this attack as N.

\noindent \textbf{Strong \quad}
We set the constraint for CW to 100 and the $\epsilon=0.2$ for PGD and APGD. We term this attack S.

\noindent \textbf{Generation}
We generate adversarial samples on five MEL datasets: Wikidata-MEL, Richpedia-MEL, WikiDiverse, WIKIPerson, and $M^3EL$. For PGD and APGD, we maximize the Cross-Entropy loss between the model logits and the ground-truth label. For CW, we minimize the sum of the $l_2$ distance of the perturbation $\delta$ and the $f$-function from the original paper~\cite{cui2024robustness}. Figure~\ref{fig: adversarial_samples} shows adversarial samples. 

\begin{figure}[!t]
    \centering
    \includegraphics[width=\columnwidth]{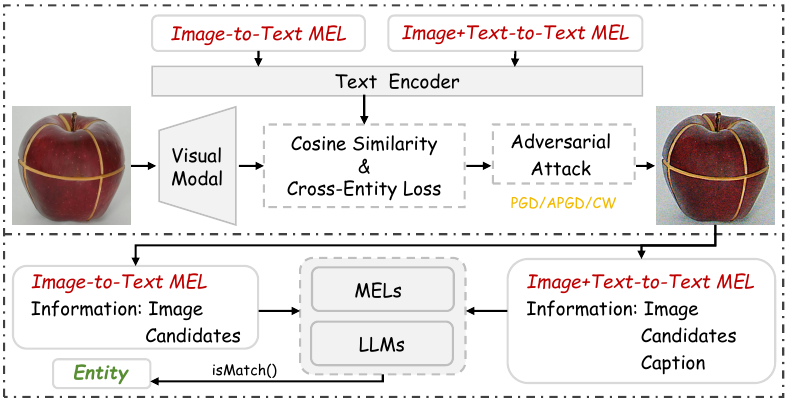}
    \caption{Overview of our procedure for attack generation and evaluation over I2T MEL and IT2T MEL.}
    \label{fig: framework}
\end{figure}

As shown in Figure~\ref{fig: framework}, we first used the text encoder to encode the text entity labels in the format of "\textit{a photo of \textless entity\textgreater}". Then, we computed the cosine similarity between the image encoding and encoded labels and used the result as the prediction logit for adversarial generation and evaluation.

\subsection{Tasks \& Evaluation}
To evaluate the impact of visual adversarial on MEL models, we consider two commonly used tasks: Image-to-Text Entity Linking and Image+Text-to-Text Entity Linking. The detailed evaluation process for the specific task is as follows.

\subsubsection{I2T-EL: Image-to-Text Entity Linking}
We systematically evaluate the robustness of I2T-EL under visual adversarial perturbations using images and corresponding candidates from five standard datasets. The evaluation covers both traditional MEL models and state-of-the-art Multimodal Large Language Models (MLLMs).

For traditional MEL methods, we first construct all candidates in the format of ``\textit{a photo of \textless candidate \textgreater}", and obtain their embeddings via a text encoder. We then compute the cosine similarity between the image embedding and each candidate embedding, and select the entity with the highest similarity score as the model prediction. The final linking accuracy is evaluated based on this prediction.

For MLLMs, the adversarial image and candidates are jointly provided as input, along with a tailored prompt: ``\textit{Which candidate best describes the main object in this image? \textbackslash n Answer with the index of the candidate.}". The linking performance is then assessed based on the model's response.

\subsubsection{Image+Text-to-Text Entity Linking}
The robustness of IT2T-EL is evaluated on five standard datasets, using images, entity descriptions, and candidates, following the same baselines as in the I2T-EL task.

For traditional MEL methods, we first obtain the embeddings of the candidates using the same procedure. The entity description is encoded using the same text encoder and fused with the image embedding through element-wise addition to form a unified multimodal representation. We then compute the cosine similarity between this fused representation and each candidate embedding, selecting the candidate with the highest similarity as the prediction to assess linking accuracy.

For MLLMs, the adversarial image, entity description, and candidates are jointly used to construct the prompt: ``\textit{Please observe the image and understand the description. Which candidate best describes the main object in the image? \textbackslash n Answer with the index of the candidate.}". 

\subsection{LLM and Retrieval-Augmented Entity Linking}
We propose a robust MEL method, LLM-RetLink (LLM and Retrieval-Augmented Entity Linking), specifically designed for complex real-world visual scenarios. The framework adopts a two-stage architecture that effectively integrates LVMs with a web-based dynamic retrieval mechanism. The goal is to significantly enhance both the robustness of the system against adversarial perturbations and the accuracy of entity linking.

In the first stage, LLM-RetLink leverages advanced vision-language models to automatically extract potential word-level entity descriptors (e.g., "apple", "fruit") from input images. These descriptors serve as semantically rich query cues for the subsequent retrieval module, thereby substantially enhancing the framework's initial perception of salient entities in the visual input.

In the second stage, we introduce a retrieval-augmented mechanism grounded in external knowledge sources. Specifically, the framework constructs sentence-level descriptions associated with candidates and dynamically queries web-based knowledge bases such as Wikidata and Wikipedia to obtain highly relevant semantic evidence based on the current context.

The retrieved content is then fed into an LLM to enhance the semantic matching between candidates and context. Unlike conventional approaches that rely on static entity descriptions, LLM-RetLink can dynamically adapt to varying contexts in open-domain settings and retrieve up-to-date, context-relevant information. This capability effectively mitigates challenges related to entity ambiguity and representation inconsistency, making the method particularly well suited for multimodal entity understanding in complex environments.

\section{Experiments}

\begin{table*}[!t]
\scalebox{0.77}{
\centering
\renewcommand{\arraystretch}{1.3}
\begin{tabular*}{\textheight}{@{\extracolsep\fill} cccccccccccccccc}
\toprule
\multirow{3}{*}{Dataset} & \multirow{3}{*}{Models} & \multicolumn{7}{c}{\textbf{Image-to-Text}} & \multicolumn{7}{c}{\textbf{Image+Text-to-Text}} \\ \cmidrule(lr){3-9} \cmidrule(lr){10-16}
                         &                         & RAW  & \multicolumn{2}{c}{PGD} & \multicolumn{2}{c}{APGD} & \multicolumn{2}{c}{CW} & RAW  & \multicolumn{2}{c}{PGD} & \multicolumn{2}{c}{APGD} & \multicolumn{2}{c}{CW} \\ \cmidrule(lr){3-3} \cmidrule(lr){4-5} \cmidrule(lr){6-7} \cmidrule(lr){8-9} \cmidrule(lr){10-10} \cmidrule(lr){11-12} \cmidrule(lr){13-14} \cmidrule(lr){15-16}
                         & & P & N & S & N & S & N & S & P & N & S & N & S & N & S         \\ \hline
\multirow{6}{*}{Wikidata-MEL} & ALIGN 	  & 30.1 & 26.8 & 22.3 & 28.5 & 27.2 & 24.2 & 21.8 &28.3 & 27.4 & 27.1 & 27.5 & 27.0 & 27.3 & 26.9  \\
                              & BLIP  	& 45.4 & 44.5 & 42.4 & 43.6 & 42.8 & 43.3 & 42.7 & 67.3 & 45.0 & 44.3 &60.1 & 55.7 & 31.4 & 29.8  \\
                              & CLIP 		& 50.1 & 31.4 & 28.3 & 32.6 & 33.1 & 24.3 & 23.2 & 66.1 & 42.8 & 31.0 & 47.5 & 45.2 & 28.7 & 26.7  \\
                              & FLAVA  	& 50.7 & 27.0 & 24.3 & 27.1 & 26.5 & 30.4 & 29.8 & 64.0 & 52.5 & 47.7 & 53.3 & 52.1 & 50.4 & 49.6  \\
                              & OWL-ViT 	& 45.3 & 45.2 & 44.9 & 45.2 & 44.1 & 45.1 & 43.8 & 49.3 & 41.0 & 35.9 & 40.8 & 39.9 & 38.1 & 36.5  \\
                              & SigLIP  	& 47.4 & 40.3 & 43.6 & 42.3 & 44.2 & 42.6 & 40.5 & 54.6 & 44.9 & 37.2 & 40.3 & 36.5 & 40.1 & 38.0  \\  \hline
\multirow{6}{*}{Richpedia-MEL}& ALIGN 	  & 20.6 & 17.0 & 12.7 & 18.4 & 16.9 & 16.1 & 12.5 &35.1 & 33.9 & 32.6 & 34.9 & 33.7 & 33.9 & 31.8  \\
                              & BLIP  	& 38.7 & 36.7 & 33.1 & 32.6 & 32.4 & 30.9 & 27.3 & 63.9 & 49.1 & 25.4 &58.3 & 55.0 & 34.1 & 32.7  \\
                              & CLIP 		& 38.8 & 21.1 & 16.4 & 21.6 & 20.0 & 15.8 & 12.7 & 60.9 & 42.3 & 34.4 & 45.9 & 43.8 & 32.3 & 28.2  \\
                              & FLAVA  	& 35.8 & 18.4 & 16.3 & 17.8 & 17.4 & 18.0 & 14.6 & 59.2 & 52.8 & 45.9 & 55.8 & 54.5 & 52.7 & 47.3  \\
                              & OWL-ViT 	& 33.2 & 19.6 & 19.2 &19.7 & 19.2 & 19.3 & 16.8 & 49.5 & 37.2 & 28.9 & 38.3 & 37.1 & 33.1 & 35.5  \\
                              & SigLIP  	& 34.5 &18.9 & 17.3 & 18.5 & 17.9 & 18.7 & 15.9 & 55.7 & 46.0 & 39.5 & 42.6 & 39.9 & 43.5 & 41.4  \\ \hline          
\multirow{6}{*}{WikiDiverse}  & ALIGN 	  & 56.1 & 55.7 & 54.2 & 53.5 & 52.8 & 55.6 & 51.4 &62.6 & 59.8 & 57.4 & 58.5 & 56.0 & 60.7 & 59.4  \\
                              & BLIP  	& 36.6 & 33.8 & 32.4 & 33.6 & 33.1 & 32.4 & 33.0 & 53.9 & 41.5 & 35.3 & 40.2 & 38.2 & 36.9 & 34.1  \\
                              & CLIP 		& 52.5 & 41.5 & 39.7 & 41.8 & 39.4 & 36.6 & 35.5 & 64.2 & 49.8 & 47.9 & 50.8 & 50.5 & 49.1 & 46.3  \\
                              & FLAVA  	& 51.0 & 39.1 & 36.9 & 40.7 & 40.0 & 36.6 & 36.0 & 67.6 & 62.7 & 62.1 & 62.0 & 61.7 & 61.5 & 58.6  \\
                              & OWL-ViT 	& 33.8 & 34.4 & 34.1 & 37.5 & 34.4 & 35.1 & 33.3 & 53.2 & 52.7 & 51.0 & 52.5 & 51.3 & 51.9 & 48.5  \\
                              & SigLIP  	& 49.7 & 36.1 & 30.4 & 40.6 & 36.5 & 36.1 & 34.8 & 60.4 & 57.3 & 56.5 & 57.2 & 56.4 & 55.7 & 52.8  \\   \hline  
\multirow{6}{*}{WIKIPerson}   & ALIGN 	  & 47.3 & 33.1 & 16.8 & 35.2 & 19.7 & 44.1 & 42.8 &55.5 & 33.8 & 21.6 & 36.4 & 21.7 & 47.9 & 47.2  \\
                              & BLIP  	& 30.1 & 35.8 & 27.7 & 33.7 & 27.6 & 29.3 & 24.4 & 53.3 & 48.2 & 35.4 & 49.6 & 48.0 & 32.5 & 29.4  \\
                              & CLIP 		& 33.7 & 21.6 & 21.1 & 21.7 & 20.9 & 19.9 & 16.7 & 45.4 & 33.3 & 29.9 & 32.0 & 30.7 & 32.4 & 27.5  \\
                              & FLAVA  	& 51.4 & 36.0 & 33.0 & 36.1 & 32.7 & 28.9 & 28.2 & 74.3 & 56.2 & 51.6 & 57.4 & 52.3 & 47.4 & 45.1  \\
                              & OWL-ViT 	& 54.1 & 53.8 & 52.4 & 53.9 & 52.3 & 54.9 & 53.3 & 68.1 & 61.4 & 58.0 & 58.3 & 56.5 & 61.7 & 60.6  \\
                              & SigLIP  	& 52.7 & 33.6 & 26.9 & 32.4 & 28.7 & 38.3 & 22.6 & 69.9 & 60.7 & 49.1 & 56.5 & 50.7 & 63.1 & 48.1  \\ \hline
\multirow{6}{*}{$M^3EL$}	  & ALIGN 	  & 30.2 & 20.7 & 16.8 & 20.2 & 18.7 & 17.8 & 15.2 &37.8 & 25.9 & 25.3 & 25.4 & 24.7 & 25.3 & 21.7  \\
                              & BLIP  	& 28.4 & 17.5 & 16.2 & 18.7 & 16.6 & 16.9 & 14.1 & 66.7 & 48.3 & 46.2 &45.5 & 39.1 & 44.3 & 39.6  \\
                              & CLIP 		& 32.7 & 19.1 & 16.9 & 19.9 & 18.1 & 17.1 & 14.6 & 68.8 & 49.7 & 45.4 & 49.3 & 42.5 & 45.7 & 41.3  \\
                              & FLAVA  	& 38.9 & 19.9 & 18.4 & 20.2 & 19.6 & 17.4 & 14.9 & 61.8 & 40.1 &37.0 & 41.5 & 39.2 & 35.9 & 31.8  \\
                              & OWL-ViT 	& 33.2 & 19.6 & 19.2 &19.7 & 19.2 & 19.3 & 16.8 & 58.4 & 26.5 & 23.4 & 27.6 & 20.3 & 23.7 & 20.2  \\
                              & SigLIP  	& 34.3 &19.8 & 18.8 & 20.1 & 19.5 & 18.6 & 15.7 & 60.2 & 33.6 & 30.8 & 29.4 & 25.2 & 28.4 & 26.5  \\ 
\bottomrule
\end{tabular*}}
\caption{Results on the robustness evaluation of different traditional MEL models across five datasets. Adversarial examples are used as input image along with contexts as input text. The numbers indicate the linking accuracy (\%). } 
\label{tab: MELs}
\end{table*}

\begin{table*}[!t]
\scalebox{0.77}{
\centering
\renewcommand{\arraystretch}{1.3}
\begin{tabular*}{\textheight}{@{\extracolsep\fill} cccccccccccccccc}
\toprule
\multirow{3}{*}{Dataset} & \multirow{3}{*}{Models} & \multicolumn{7}{c}{\textbf{Image-to-Text}} & \multicolumn{7}{c}{\textbf{Image+Text-to-Text}} \\ \cmidrule(lr){3-9} \cmidrule(lr){10-16}
                         &                         & RAW  & \multicolumn{2}{c}{PGD} & \multicolumn{2}{c}{APGD} & \multicolumn{2}{c}{CW} & RAW  & \multicolumn{2}{c}{PGD} & \multicolumn{2}{c}{APGD} & \multicolumn{2}{c}{CW} \\ \cmidrule(lr){3-3} \cmidrule(lr){4-5} \cmidrule(lr){6-7} \cmidrule(lr){8-9} \cmidrule(lr){10-10} \cmidrule(lr){11-12} \cmidrule(lr){13-14} \cmidrule(lr){15-16}
                         & & P & N & S & N & S & N & S & P & N & S & N & S & N & S         \\ \hline
\multirow{6}{*}{Wikidata-MEL} & ALIGN* 	  & 52.2 & 50.7 & 49.4 & 50.7 & 47.3 & 51.1 & 37.6 & 47.3 & 46.6 & 46.3 & 46.3 & 46.1 & 46.7 & 44.1  \\
                              & BLIP*  	& 65.2 & 64.9 & 64.3 & 65.0 & 64.7 & 55.8 & 55.4 & 97.4 & 62.2 & 61.8 & 77.1 & 72.9 & 49.4 & 48.2 \\
                              & CLIP*  	& 69.4 & 43.8 & 41.0 & 44.1 & 43.1 & 40.7 & 38.9 & 94.7 & 62.6 & 52.4 & 65.2 & 63.3 & 50.9 & 49.7  \\
                              & FLAVA*  	& 74.2 & 63.2 & 62.2 & 64.2 & 64.0 & 68.0 & 67.5 & 94.0 & 63.9 & 58.8 & 64.8 & 64.4 & 63.2 & 62.7  \\
                              & OWL-ViT*  & 60.5 & 62.4 & 61.1 & 63.4 & 62.5 & 61.7 & 60.8 & 76.6 & 67.4 & 62.6 & 67.7 & 66.6 & 63.3 & 60.3  \\
                              & SigLIP*  	& 66.0 & 62.1 & 60.1& 64.7 & 60.9 & 69.6 & 67.7 & 79.4 & 60.3 & 50.4 & 58.8 & 55.8 & 50.6 & 49.4  \\ \hline
\multirow{6}{*}{Richpedia-MEL}& ALIGN* 	  & 49.1 & 40.6 & 39.3 & 41.8 & 38.4 & 40.7 & 36.2 & 46.4 & 45.0 & 44.8 & 45.2 & 44.1 & 44.9 & 41.8  \\
                              & BLIP*  	& 69.7 & 59.8 & 53.1 & 57.3 & 49.6 & 47.7 & 36.1 & 93.8 & 68.2 & 63.9 & 77.4 & 73.7 & 52.5 & 53.6  \\
                              & CLIP*  	& 68.8 & 65.3 & 61.5 & 58.2 & 54.8 & 48.4 & 45.8 & 90.8 & 51.9 & 52.1 & 54.2 & 52.0 & 40.3 & 38.2  \\
                              & FLAVA*  	& 66.9 & 61.6 & 60.9 & 59.0 & 51.7 & 53.8 & 46.6 & 88.7 & 56.0 & 50.8 & 57.9 & 57.2 & 53.6 & 52.2  \\
                              & OWL-ViT* 	& 69.2 & 63.1 & 58.3 & 61.4 & 54.2 & 51.8 & 43.7 & 73.1 & 60.4 & 60.9 & 51.9 & 59.0 & 59.1 & 55.5  \\
                              & SigLIP*  	& 48.5 & 20.6 & 20.1 & 22.6 & 21.5 & 20.2 & 16.7 & 78.3 & 54.5 & 55.6 & 55.5 & 53.7 & 52.2 & 50.3  \\  \hline
\multirow{6}{*}{WikiDiverse}  & ALIGN* 	  & 70.6 & 57.8 & 46.3 & 52.7 & 51.0 & 47.3 & 46.9 & 79.0 & 62.9 & 61.8 & 63.5 & 61.4 & 64.3 & 62.8  \\
                              & BLIP*  	& 51.9 & 49.2 & 48.5 & 47.8 & 47.3 & 46.9 & 46.1 & 64.0 & 52.2 & 47.0 & 52.2 & 49.7 & 48.4 & 45.5  \\
                              & CLIP*  	& 64.1 & 57.9 & 56.8 & 58.2 & 57.5 & 58.2 & 57.1 & 76.1 & 63.2 & 62.9 & 64.7 & 64.1 & 60.3 & 59.8  \\
                              & FLAVA*  	& 67.4 & 64.5 & 63.5 & 64.2 & 63.8 & 64.7 & 63.8 & 80.2 & 73.8 & 70.6 & 74.1 & 73.4 & 70.3 & 69.5  \\
                              & OWL-ViT* 	& 59.8 & 59.6 & 59.4 & 59.9 & 59.7 & 59.5 & 58.1 & 67.5 & 66.7 & 64.2 & 66.5 & 66.0 & 66.3 & 61.2  \\
                              & SigLIP*  	& 62.2 & 55.1 & 53.2 & 56.6 & 54.9 & 52.6 & 50.3 & 74.3 & 70.2 & 67.4 & 71.3 & 69.8 & 67.6 & 66.1  \\ \hline
\multirow{6}{*}{WIKIPerson}   & ALIGN* 	  & 89.5 & 59.7 & 42.7 & 58.2 & 42.8 & 49.7 & 45.2 & 94.4 & 63.8 & 45.9 & 67.2 & 44.7 & 64.3 & 51.5  \\
                              & BLIP*  	& 56.1 & 52.1 & 50.4 & 55.8 & 52.4 & 52.3 & 47.1 & 80.2 & 62.5 & 61.2 & 76.5 & 75.8 & 64.6 & 57.1  \\
                              & CLIP*  	& 82.7 & 44.5 & 44.0 & 43.6 & 42.1 & 34.5 & 32.9 & 74.7 & 52.1 & 49.8 & 50.1 & 49.5 & 36.6 & 35.6  \\
                              & FLAVA*  	& 71.5 & 66.5 & 65.8 & 66.3 & 66.1 & 64.6 & 59.7 & 92.7 & 91.8 & 87.8 & 93.4 & 92.0 & 90.4 & 80.6  \\
                              & OWL-ViT* 	& 87.6 & 87.6 & 86.3 & 87.7 & 84.2 & 87.5 & 85.6 & 78.8 & 58.6 & 48.4 & 56.7 & 42.8 & 59.9 & 43.1  \\
                              & SigLIP*  	& 66.4 & 54.8 & 53.2 & 56.5 & 54.2 & 52.4 & 51.3 & 89.6 & 88.7 & 75.6 & 89.4 & 76.4 & 84.4 & 74.9  \\  \hline
\multirow{6}{*}{$M^3EL$}	  & ALIGN* 	  & 44.9 & 21.9 & 19.3 & 24.0 & 23.4 & 19.8 & 16.3 & 48.7 & 29.0 & 28.1 & 28.1 & 27.6 & 28.0 & 24.3  \\
                              & BLIP*  	& 30.0 & 29.4 & 27.0 & 19.5 & 17.4 & 17.5 & 17.2 & 87.1 & 49.7 & 48.6 & 48.5 & 40.1 & 47.2 & 41.4  \\
                              & CLIP*  	& 36.3 & 20.8 & 18.4 & 20.0 & 19.7 & 19.4 & 18.1 & 88.6 & 52.6 & 48.7 & 51.8 & 50.5 & 43.8 & 40.7  \\
                              & FLAVA*  	& 44.8 & 20.4 & 20.1 & 20.4 & 20.3 & 19.8 & 16.5 & 85.0 & 45.7 & 42.1 & 47.5 & 46.2 & 45.6 & 37.4  \\
                              & OWL-ViT*  & 36.7 & 20.9 & 20.3 & 23.3 & 23.2 & 20.4 & 17.0 & 69.5 & 31.6 & 28.3 & 32.4 & 32.2 & 28.7 & 26.1  \\
                              & SigLIP*  	& 38.5 & 20.6 & 20.1 & 22.6 & 21.5 & 20.2 & 16.7 & 72.4 & 38.5 & 33.2 & 39.9 & 37.3 & 31.5 & 29.3  \\  
\bottomrule
\end{tabular*}}
\caption{Results on the robustness evaluation of different traditional MEL models across five datasets. \textit{Model*} indicates the use of the LLM-RetLink method. The numbers indicate the linking accuracy (\%).} 
\label{tab: MELsDriven}
\end{table*}

\subsection{Experimental Setups}
\subsubsection{Datasets}

\begin{itemize}
\item \textbf{Wikidata-MEL}~\cite{ZhouWLXW21} contains over 18K multimodal samples extracted from Wikipedia\footnote{https://en.wikipedia.org} and Wikidata\footnote{https://www.wikidata.org}, the majority of the entity types in the dataset are Person.

\item \textbf{Richpedia-MEL}~\cite{ZhouWLXW21} is jointly constructed from Richpedia~\cite{Richpedia} and Wikipedia, containing over 17K multimodal samples.

\item \textbf{WikiDiverse}~\cite{wang2022wikidiverse} is a manually annotated MEL dataset consisting of diverse contextual topics and 13 entity types derived from Wikinews~\footnote{https://www.wikinews.org/}. 

\item \textbf{WIKIPerson}~\cite{WikiPerson} constructed based on Wikipedia for the Visual Named Entity Linking task. It contains approximately 16 million entities across 13 categories, and covers a wide range of topics.

\item \textbf{$M^3EL$}~\cite{WF-M3EL} is a large-scale and high-quality MEL dataset, manually labeled and constructed based on Kaggle\footnote{https://www.kaggle.com}, DBpedia\footnote{https://www.dbpedia.org}, Wikipedia, and Wikidata. It includes 79,625 instances, covering 9 diverse multimodal tasks and 5 distinct topics (e.g., Movie, General Knowledge, Person, Book, Sport).
\end{itemize}

\subsubsection{Baselines}
We compared our method with recent state-of-the-art methods, which are divided into two groups: traditional MEL methods (ALIGN, BLIP, CLIP, FLAVA, OWL-ViT, SigLIP) and multimodal large language models (LLaVA, Qwen2.5-VL, MiniGPT-4). The following are the details of the baselines.

\begin{itemize}
\item \textbf{ALIGN}~\cite{align} features a dual-encoder architecture with EfficientNet as its vision encoder and BERT as its text encoder, and learns to align visual and text representations with contrastive learning.
\item \textbf{BLIP}~\cite{blip2} leverages frozen pre-trained image encoders and large language models (LLMs) by training a lightweight, 12-layer Transformer encoder in between them, fusing visual, textual of mentions
\item \textbf{CLIP}~\cite{clip} implements separate Transformer encoders for each modality, enabling joint learning of visual-textual feature correlations.
\item \textbf{FLAVA}~\cite{flava} is a single unified foundation model that can work across vision, language as well as vision-and-language multimodal tasks.
\item \textbf{OWL-ViT}~\cite{OWL-ViT} can be used to query an image with one or multiple text queries to search for and detect target objects described in text.
\item \textbf{SigLIP}~\cite{siglip} operates solely on image-text pairs and does not require a global view of the pairwise similarities for normalization.
\item \textbf{LLaVA}~\cite{Llava} is an open-source chatbot trained by fine-tuning LlamA/Vicuna on GPT-generated multimodal instruction-following data.
\item \textbf{Qwen2.5-VL}~\cite{qwen2} is a multimodal vision-language model with strong visual recognition capabilities, capable of identifying common objects, charts, layouts, and other elements.
\item \textbf{MiniGPT-4}~\cite{MiniGPT} is a large multimodal model based on deep learning, capable of processing multimodal data.
\end{itemize}

\subsubsection{Implementation Details}
Experiments were run on A100 GPU (40GB), averaged over 3 runs. The temperature of MLLMS set to 0.75 (for consistency in fixed output formats) and other parameters remaining at their default settings. We use Accuracy ( precision \textbf{@}1) to evaluate MEL effectiveness in all experiments.

\begin{equation}
    \text{Acc} = \frac{1}{N} \sum_{i=1}^{N} \mathbb{I}(C_i \leftrightarrow E)
\end{equation}

where $M_i$ is the candidate, $E$ is the correct entity, $N$ is total instances, and the indicator function $\mathbb{I}(C_i \leftrightarrow E)$ denotes that $C_i$ corresponds to entity $E$.


\begin{table*}[!t]
\scalebox{0.77}{
\centering
\renewcommand{\arraystretch}{1.3}
\begin{tabular*}{\textheight}{@{\extracolsep\fill} cccccccccccccccc}
\toprule
\multirow{3}{*}{Dataset} & \multirow{3}{*}{Models} & \multicolumn{7}{c}{\textbf{Image-to-Text}} & \multicolumn{7}{c}{\textbf{Image+Text-to-Text}} \\ \cmidrule(lr){3-9} \cmidrule(lr){10-16}
                         &                         & RAW  & \multicolumn{2}{c}{PGD} & \multicolumn{2}{c}{APGD} & \multicolumn{2}{c}{CW} & RAW  & \multicolumn{2}{c}{PGD} & \multicolumn{2}{c}{APGD} & \multicolumn{2}{c}{CW} \\ \cmidrule(lr){3-3} \cmidrule(lr){4-5} \cmidrule(lr){6-7} \cmidrule(lr){8-9} \cmidrule(lr){10-10} \cmidrule(lr){11-12} \cmidrule(lr){13-14} \cmidrule(lr){15-16}
                         & & P & N & S & N & S & N & S & P & N & S & N & S & N & S         \\ \hline
\multirow{3}{*}{Wikidata-MEL} & LLaVA	  & 61.7 & 45.1 & 42.6 & 42.3 & 41.2 & 39.6 & 37.4 & 66.6 & 49.2 & 45.7 & 46.7 & 42.4 & 44.5 & 43.9  \\
                              & LLaVA*	& 74.7 & 59.3 & 57.5 & 58.6 & 56.4 & 55.2 & 44.6 & 86.9 & 63.6 & 57.8 & 71.7 & 64.6 & 56.6 & 55.0  \\
                              & Qwen2.5-VL& 63.5 & 49.5 & 48.0 & 45.3 & 35.8 & 34.8 & 27.7 & 78.8 & 58.3 & 49.4 & 56.4 & 46.0 & 45.3 & 44.8  \\ 
                              & Qwen2.5-VL*& 78.3 & 69.2 & 67.2 & 65.4 & 56.7 & 62.1 & 56.3 & 88.1 & 70.6 & 69.9 & 66.9 & 61.1 & 65.6 & 57.1  \\
                              & MiniGPT-4	& 75.0 & 64.4 & 60.3 & 62.7 & 51.9 & 54.4 & 41.5 & 91.1 & 79.4 & 71.8 & 77.3 & 76.2 & 66.2 & 52.2  \\ 
                              & MiniGPT-4*& 80.5 & 67.4 & 61.8 & 69.2 & 63.5 & 52.3 & 48.4 & 93.9 & 89.2 & 81.5 & 85.1 & 78.6 & 71.3 & 66.7  \\ \hline
\multirow{3}{*}{Richpedia-MEL}& LLaVA	  & 69.0 & 65.7 & 60.4 & 63.2 & 59.3 & 59.4 & 57.8 & 83.5 & 82.6 & 80.3 & 79.7 & 76.1 &67.2 & 63.6  \\
                              & LLaVA*	& 87.1 & 85.4 & 80.8 & 78.0 & 74.8 & 64.2 & 61.6 & 91.2 & 87.8 & 78.5 & 83.7 & 77.4 & 69.4 & 67.0  \\
                              & Qwen2.5-VL& 70.1 & 68.4 & 66.9 & 65.4 & 61.2 & 59.7 & 58.1 & 85.2 & 81.6 & 79.4 & 80.7 & 76.5 & 69.2 & 68.6  \\ 
                              & Qwen2.5-VL*& 93.1 & 86.1 & 81.7 & 89.2 & 78.3 & 71.9 & 66.3 & 96.5 & 95.5 & 86.6 & 87.2 & 80.9 & 76.6 & 73.5  \\
                              & MiniGPT-4	& 74.6 & 70.5 & 67.4 & 66.5 & 62.8 & 60.6 & 58.8 & 90.9 & 88.1 & 81.7 & 85.1 & 80.7 & 77.1 & 71.9  \\ 
                              & MiniGPT-4*& 95.9 & 89.3 & 87.1 & 91.7 & 84.1 & 76.2 & 69.7 & 97.1 & 96.4 & 89.3 & 93.8 & 86.1 & 81.0 & 75.2  \\ \hline
\multirow{3}{*}{WikiDiverse}  & LLaVA	  & 71.7 &66.2 & 61.5 & 65.4 & 62.1 & 56.9 & 51.6 & 78.1 & 76.9 & 71.3 & 76.2 & 72.5 & 69.3 & 67.5  \\
                              & LLaVA*	& 84.5 & 74.8 & 63.7 & 70.3 & 65.8 & 57.3 & 52.4 & 86.6 & 77.8 & 72.2 & 77.4 & 76.8 & 71.8 & 69.8  \\
                              & Qwen2.5-VL& 73.8 & 70.6 & 67.4 & 63.9 & 61.6 & 57.1 & 54.2 & 83.8 & 78.3 & 73.5 & 77.4 & 76.2 & 69.5 & 68.6  \\ 
                              & Qwen2.5-VL*& 87.1 & 77.2 & 74.1 & 79.9 & 67.9 & 63.6 & 60.9 & 89.9 & 86.5 & 81.2 & 89.8 & 77.5 & 79.2 & 76.6  \\
                              & MiniGPT-4	& 77.3 & 73.5 & 68.3 & 64.7 & 60.4 & 57.1 & 54.7 & 85.2 & 80.6 & 77.1 & 78.6 & 77.3 & 69.8 & 69.4  \\ 
                              & MiniGPT-4*& 89.7 & 78.0 & 75.4 & 80.6 & 76.2 & 68.4 & 62.5 & 96.8 & 93.1 & 89.6 & 95.3 & 91.1 & 83.4 & 78.3  \\ \hline
\multirow{3}{*}{WIKIPerson}   & LLaVA	  & 61.7 & 45.1 & 42.6 & 42.3 & 41.2 & 39.6 & 37.4 & 69.9 & 45.8 & 47.4 & 43.5 & 42.3 & 43.4 & 41.8  \\
                              & LLaVA*	& 78.9 & 67.8 & 63.5 & 63.8 & 59.1 & 64.1 & 58.7 & 80.1 & 71.2 & 70.3 & 69.7 & 65.8 & 67.6 & 63.2  \\
                              & Qwen-VL	& 63.5 & 49.5 & 48.4 & 45.3 & 35.8 & 34.8 & 27.7 & 75.2 & 61.2 & 56.8 & 55.9 & 45.6 & 47.3 & 38.5  \\ 
                              & Qwen-VL* 	& 83.3 & 73.9 & 69.9 & 79.7 & 67.6 & 68.6 & 59.2 & 89.5 & 80.6 & 79.1 & 85.6 & 76.4 & 71.0 & 69.6  \\
                              & MiniGPT-4	& 75.0 & 64.4 & 60.3 & 62.7 & 51.9 & 54.4 & 41.5 & 84.4 & 72.7 & 67.7 & 64.5 & 68.8 & 56.2 & 51.7  \\ 
                              & MiniGPT-4*& 90.9 & 86.1 & 84.2 & 85.8 & 79.4 & 71.9 & 63.6 & 92.2 & 87.3 & 80.3 & 90.4 & 88.3 & 81.5 & 79.5  \\ \hline
\multirow{3}{*}{$M^3EL$}      & LLaVA	  & 40.8 & 37.1 & 36.2 & 36.5 & 31.8 & 33.2 & 22.4 & 46.6 & 45.3 & 44.7 & 41.4 & 36.5 & 37.1 & 30.4  \\
                              & LLaVA*	& 55.7 & 46.4 & 44.9 & 41.3 & 37.8 & 35.5 & 32.1 & 72.8 & 66.7 & 64.2 & 70.2 & 65.6 & 56.2 & 51.3  \\
                              & Qwen2.5-VL& 45.9 & 40.8 & 35.4 & 38.5 & 34.2 & 37.4 & 36.6 & 68.2 & 57.6 & 47.9 & 50.3 & 45.4 & 49.2 & 41.8  \\ 
                              & Qwen2.5-VL*& 64.4 & 51.2 & 47.1 & 58.9 & 54.8 & 57.8 & 48.2 & 84.6 & 76.5 & 69.1 & 86.0 & 80.7 & 73.6 & 69.1  \\
                              & MiniGPT-4	& 47.5 & 41.9 & 36.9 & 39.9 & 36.5 & 38.2 & 36.9 & 80.3 & 62.6 & 54.9 & 69.1 & 55.5 & 56.5 & 52.9  \\
                              & MiniGPT-4*& 79.3 & 62.2 & 57.1 & 60.6 & 58.2 & 58.4 & 52.7 & 93.2 & 82.1 & 78.2 & 77.9 & 74.3 & 79.7 & 75.3  \\
\bottomrule
\end{tabular*}}
\caption{Results on the robustness evaluation of different MLLMs on the MEL task across across five datasets. \textit{Model*} indicates the use of the LLM-RetLink method. The numbers indicate the linking accuracy (\%).} 
\label{tab: LLMs}
\end{table*}

\subsection{Experimental Results}
\subsubsection{Robustness Evaluation of Traditional MEL Models}
\quad As shown in Table~\ref{tab: MELs}, all models exhibit varying degrees of performance degradation under adversarial attacks, highlighting the vulnerability of MEL tasks in real-world multimodal adversarial settings. 

In the Image-to-Text task, ALIGN, OWL-ViT, and BLIP exhibit notable robustness under three types of adversarial attacks, with average accuracy drops of 5.1\%, 6.1\%, and 6.4\%, respectively. For example, BLIP achieves 45.4\% RAW accuracy on the Wikidata-MEL dataset, with performance degradation limited to 2–3\% under PGD and APGD attacks, indicating its resilience to noise interference. Notably, OWL-ViT exhibits strong stability on the WikiDiverse dataset, particularly under Normal attacks, where performance fluctuations are minimal, suggesting that its architectural design may be better suited for handling diverse data. In contrast, SigLIP, CLIP, and FLAVA show significant performance degradation under adversarial attacks, with average accuracy drops of 15.6\%, 17.3\%, and 18.3\%, respectively, revealing their vulnerability. 

In the Image+Text-to-Text task, the introduction of multimodal inputs has a pronounced impact on model robustness. FLAVA stands out with exceptional stability across all datasets, maintaining an accuracy of 49.6\% even under CW-S attacks on Wikidata-MEL, demonstrating that its cross-modal fusion mechanism effectively mitigates adversarial interference. However, BLIP exhibits a polarized performance: while excelling in RAW linking, its accuracy can plummet by over 30\% under Strong attacks. In particular, ALIGN performs better in $M^3EL$, with only a 4.2\% drop under CW attacks, suggesting that data diversity may enhance model robustness. 

Overall, traditional MEL methods tend to suffer from insufficient robustness against adversarial interference. However, models with stronger multimodal alignment capabilities and more complex architectural designs can partially mitigate performance degradation. These findings indicate that improving cross-modal robustness remains a critical challenge for the practical deployment of MEL.

\subsubsection{Effectiveness of the LLM-RetLink Method} 
As shown in Table~\ref{tab: MELsDriven} and Table~\ref{tab: LLMs}, we conduct a comparative study on both traditional MEL models and MLLMs for the MEL task, and their respective versions enhanced with the LLM-RetLink method. The results demonstrate significant improvements in both performance and robustness.

Results for traditional MEL models are shown in Table~\ref{tab: MELsDriven}. In the Image-to-Text task, our method improves the average RAW accuracy by 21.3\% (e.g., FLAVA increases from 50.7\% to 74.2\%) and significantly enhances adversarial robustness (e.g., CLIP's accuracy under APGD-S attacks on the Richpedia-MEL rises from 12.7\% to 54.8\%). In the Image+Text-to-Text task, the method effectively reduces performance fluctuations in multimodal settings. Notably, on the WIKIPerson, ALIGN* improves accuracy under CW-N attacks from 21.6\% to 45.9\%, while also narrowing the performance gap across different attack strengths by 23.8\%. These results confirm that improving entity descriptive sentences substantially improves the robustness and stability of multimodal models.

Results of MLLMs on the MEL task in Table~\ref{tab: LLMs}. In the Image-to-Text task, the three models achieved an average improvement of 14.8\% in RAW accuracy, with Qwen2.5-VL* standing out on the Wikidata-MEL dataset, increasing from 63.5\% to 78.3\%. Furthermore, robustness against adversarial attacks was significantly enhanced, as demonstrated by MiniGPT-4* on the Richpedia-MEL dataset, where the accuracy under APGD-S attacks improved by 21.3\%. In the Image+Text-to-Text task, LLM-RetLink exhibited stronger cross-modal protection capabilities. For instance, LLaVA* improved its accuracy under CW-N attacks on WIKIPerson from 47.4\% to 70.3\%, while narrowing the performance gap across different attack intensities by 15.2\%. The performance of Qwen-VL* on WikiDiverse, where the RAW accuracy improves by 13.3\%, and the performance drop under Strong attacks is mitigated from 19.6\% to 12.2\%. Collectively, these improvements verify that LLM-RetLink effectively enhances the robustness of models in complex adversarial scenarios.

\subsubsection{Robustness Evaluation of MLLMs on the MEL Task}
Table~\ref{tab: LLMs} shows that MLLMs exhibit notable variations in robustness across different datasets and attack scenarios.

In the Image-to-Text task, LLaVA demonstrates stable performance on the Wikidata-MEL and Richpedia-MEL datasets (e.g., only a 3.3\% drop under PGD-N attacks on Richpedia-MEL). Qwen2.5-VL exhibits significant fluctuations in complex scenarios, such as an 11.7\% drop under APGD-S attacks on $M^3EL$. MiniGPT-4 achieves the best overall performance, maintaining 58.8\% accuracy under CW-S attacks on Richpedia-MEL.

In the Image+Text-to-Text task, LLaVA reveals shortcomings in cross-modal defense (a 20.1\% drop under CW-S attacks on WikiDiverse). Qwen2.5-VL shows instability in handling the text modality, with a sharp 35.8\% drop under CW-S attacks on WIKIPerson. MiniGPT-4 again demonstrates the strongest overall robustness, with only an 11\% drop under APGD-S attacks on $M^3EL$. However, all models experience more than a 36\% performance drop under CW-S attacks on WIKIPerson.

\section{Conclusion}
This study systematically reveals the vulnerability of multimodal entity linking (MEL) models to visual adversarial attacks, addressing a previously overlooked gap in robustness research within this domain. Through a comprehensive evaluation of mainstream MEL models on Image-to-Text (I2T) and Image+Text-to-Text (IT2T) tasks, we clearly demonstrate the prevalent lack of anti-interference capability in current models, while also uncovering the positive role of contextual semantic information in mitigating adversarial perturbations. Based on this insight, we propose the LLM-RetLink method, which adopts a two-stage strategy: initially extracting entity descriptions using large vision models and then dynamically generating candidate descriptions via web-based retrieval, achieves an accuracy improvement of 0.4\%-35.7\% on five datasets. Furthermore, the construction and public release of the first MEL adversarial example dataset provides valuable resources to support future research in related fields.

\bibliography{aaai2026}

\end{document}